\documentclass[pss]{wiley2sp} 
\usepackage{amsmath}
\usepackage{bm}   
\usepackage{graphicx}
\usepackage{dcolumn}
\usepackage{amsfonts}
\usepackage{amssymb}
\usepackage{amsmath}
\usepackage{placeins}
\usepackage{color}
\usepackage{hyperref}
\usepackage{multirow}
\usepackage{mathtools}
\usepackage{cuted}
\usepackage{comment}

\begin{document}

\title{Heterostructures of graphene and topological insulators Bi$_2$Se$_3$, Bi$_2$Te$_3$, and Sb$_2$Te$_3$}

\author{%
  Klaus Zollner\textsuperscript{\Ast,\textsf{\bfseries 1}},
  Jaroslav Fabian\textsuperscript{\textsf{\bfseries 1}}}

\mail{\textsf{klaus.zollner@physik.uni-regensburg.de}}

\institute{%
  \textsuperscript{1}\,Institute for Theoretical Physics, University of Regensburg, 93053 Regensburg, Germany}


\abstract{\bf%
Prototypical three-dimensional topological insulators of the Bi$_2$Se$_3$ family provide a beautiful example of the appearance of the surface states inside the bulk band gap caused by spin-orbit coupling-induced topology. The surface states are protected against back scattering by time reversal symmetry, and exhibit spin-momentum locking whereby the electron spin is polarized perpendicular to the momentum, typically in the plane of the surface. On the other hand, graphene is a prototypical two-dimensional material, with negligible spin-orbit coupling. When graphene is placed on the surface of a topological insulator, giant spin-orbit coupling is induced by the proximity effect, enabling interesting novel electronic properties of its Dirac electrons. We present a detailed theoretical study of the proximity effects of monolayer graphene and topological insulators Bi$_2$Se$_3$, Bi$_2$Te$_3$, and Sb$_2$Te$_3$, and elucidate the appearance of the qualitatively new spin-orbit splittings well described by a phenomenological Hamiltonian, by analyzing the orbital decomposition of the involved band structures. This should be useful for building microscopic models of the proximity effects between the surfaces of the topological insulators and graphene.}

\maketitle   

\section{Introduction}
The three-dimensional topological 
insulators \cite{Zhang2009:NP} Bi$_2$Se$_3$, Bi$_2$Te$_3$, and Sb$_2$Te$_3$, are prototypical bulk materials demonstrating topological surface states, with a potential for practical applications, such as photodetectors and transistors \cite{Tian2017:Mat}. 
The surface states feature Dirac electrons whose spins are locked to the momentum \cite{Hsieh2009:Nat}, providing topological protection 
against back-scattering \cite{Hasan2010:RMP,Zhang2009:PRL} and making
the states highly conducting \cite{Koirala2015:NL}. 
These materials consist of quintuple layers (QLs) of alternating Bi/Sb and Se/Te atoms, where weak van der Waals forces hold the individual QLs together. 
The minimum number of QLs, such that topologically protected surface states emerge, is about 5--6, as demonstrated by angle resolved photoemission spectroscopy (ARPES) \cite{Zhang2010:NP} and 
first-principles calculations \cite{Liu2010:PRB,Yazyev2010:PRL,Park2010:PRL}.
When these topological insulators are too thin, top and bottom surface state wave functions hybridize through the bulk, and a finite gap  emerges in the Dirac spectrum.

Another prototypical material that hosts Dirac electrons is two-dimensional graphene \cite{Gmitra2009:PRB,Geim2007:NM}, which is a single layer of carbon atoms arranged in a honeycomb lattice.  
In contrast to the topological insulators, the Dirac states in graphene are not topologically protected. 
However, due to the two-dimensional nature of graphene, one can easily manipulate its electronic states via so called proximity effects \cite{Zutic2019:MT}. 
Within van der Waals heterostructures \cite{Geim2013:Nat,Novoselov2016:Sci,Duong2017:ACS} 
with other two dimensional materials, one can induce magnetism, as well as strong spin-orbit coupling (SOC) in graphene \cite{Gmitra2015:PRB,Gmitra2016:PRB,Zollner2017:NJP,Zollner2019:PRB2}.
Combining proximity-induced exchange and SOC in graphene can, under the right conditions, also lead to topologically protected edge states \cite{Frank2018:PRL,Hogl2020:PRL}, which could be important for novel spintronics applications. 

\begin{figure*}[!htb]
 \includegraphics[width=.99\textwidth]{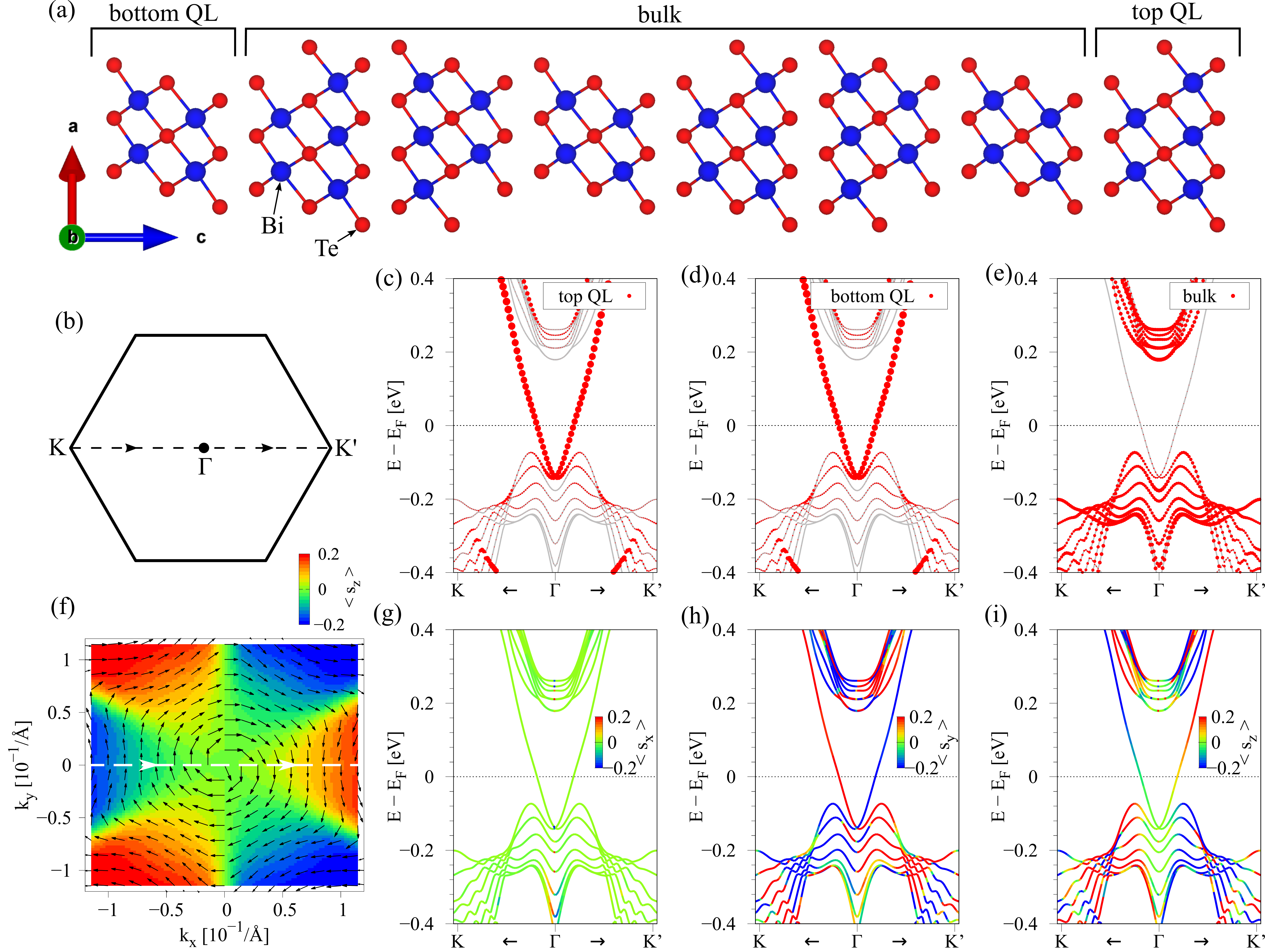}
 \caption{Geometry and calculated band structure of Bi$_2$Te$_3$. (a)
 Side view of 8 QLs of Bi$_2$Te$_3$ with definitions of the top and bottom QL and the bulk-like part of the geometry. (b) The first Brillouin Zone of the hexagonal unit cell, defining the $k$-path for the band structure. (c-e) The calculated band structure projected onto the three different parts (top QL, bottom QL, bulk) in the geometry, as defined in (a). (f) The spin-orbit field of top QL Dirac bands around the $\Gamma$ point. The color corresponds to the $s_z$ expectation value, while arrows represent the in-plane spin components. (g-i) The same as (c-e), where the color corresponds to the $s_x$, $s_y$, and $s_z$ spin expectation value, respectively. 
 }\label{Fig:Bi2Te3_geometry_bands}
\end{figure*}

There have already been numerous studies considering graphene/topological insulator bilayers \cite{Song2018:NL,Khokhriakov2018:SA,Jafarpisheh2018:PRB,Zhang2014:PRL,Zalic2017:PRB,Steinberg2015:PRB,Song2010:APL,Dang2010:NL,Lee2015:ACS,Kou2015:C,Kou2013:NL,Rodriguez2017:PRB,Cao2016:2DM,Qiao2015:ACS,Rajput2016:ACS,Vaklinova2016:NL,Popov2014:PRB}, in which two different kinds of Dirac electrons are simultaneously present.
More specifically, it is possible  to grow high-quality topological insulators such as Bi$_2$Se$_3$ epitaxially layer-by-layer on graphene, with small defect density \cite{Song2010:APL,Dang2010:NL}. Such 
heterostructures can actually be used as broadband
photodetectors \cite{Qiao2015:ACS}. Relevant to our work are spin properties of graphene/topological insulator slabs. It has been argued that these structures can still exhibit quantum spin Hall states \cite{Lee2015:ACS,Cao2016:2DM}, and that the
type and magnitude of proximity SOC in graphene can be tuned by the twist angle \cite{Zhang2014:PRL,Song2018:NL}. Finally, spin transport experiments have demonstrated spin-to-charge conversion in graphene on (Bi$_{0.15}$Sb$_{0.85}$)$_2$Te$_3$
 \cite{Khokhriakov2019:arxiv} and on Bi$_2$Te$_2$Se \cite{Vaklinova2016:NL}.

Similar to transition-metal dichalcogenides \cite{Gmitra2015:PRB,Gmitra2016:PRB}, topological insulators strongly enhance the negligible intrinsic SOC of the graphene Dirac states from 10~$\mu$eV \cite{Gmitra2009:PRB,Abdelouahed2010:PRB,Sichau2019:PRL}, by two orders of magnitude to about 1~meV 
\cite{Song2018:NL,Jafarpisheh2018:PRB,Jin2013:PRB}. 
The proximity-induced SOC in graphene is giant, drastically reducing spin relaxation times, and of valley-Zeeman type, leading to giant spin relaxation anisotropies \cite{Song2018:NL,Cummings2017:PRL}. 
Such graphene/topological insulator bilayers are ideal for the interaction of topological surface states, with in-plane spin-momentum locking, and the proximity induced spin-orbit fields in graphene.

In this manuscript,  
we first review the properties of Bi$_2$Te$_3$ as a representative example of the three dimensional topological insulator family. Our first-principles results, considering 8 layers of Bi$_2$Te$_3$ where the Dirac surface states have already formed, are consistent with literature. We include this background information to set the stage for the discussion of main results. Since we will be
interested in the effects of an external (transverse) electric field, we also investigated 
the gate effect on the degenerate surface 
states in the Bi$_2$Te$_3$ slab, which exhibit spin-momentum locking.  Indeed, we show
that these topological states can be efficiently separated in energy by an applied electric field. 
The splitting, $\Delta E$, of the surface states, increases linearly with the slope of about
6.5~meV per mV/nm with the applied field. This 
DFT prediction agrees well with a simple 
estimate based on electrostatics.

In the second part, we consider graphene/topological insulator bilayers where we are interested in the proximity-induced SOC in graphene.
We quantify the magnitude and type of induced SOC by fitting a symmetry-derived model Hamiltonian to the low energy bands of graphene. 
The proximity-induced SOC in graphene is similar, but with variations in magnitude ($0.1$--$1$~meV), for the considered topological insulators Bi$_2$Se$_3$, Bi$_2$Te$_2$Se, and Sb$_2$Te$_3$. Moreover, the charge transfer between materials and the resulting doping level of graphene can be significantly different for different topological insulators, ranging from 0 to 350~meV in terms of the Fermi energy. 
When the Dirac points of both materials are located near the Fermi level, as the case of Bi$_2$Te$_2$Se indicates, the simultaneous study of two very different spin-orbit fields is possible. 
Motivated by the recent spin-charge conversion experiments in graphene on (Bi$_{0.15}$Sb$_{0.85}$)$_2$Te$_3$ \cite{Khokhriakov2019:arxiv}, we extensively discuss the case of graphene/Sb$_2$Te$_3$,  
including spin-orbit fields, and the gate tunability of proximity-induced SOC and the doping level. 
We find a giant electric field tunability of Rashba and intrinsic SOC, in magnitude and sign, which is important to interpret the above experimental data.

\section{Monolayer graphene in proximity to Bi$_2$Se$_3$ and Bi$_2$Te$_3$}

\subsection{Topological band structure of Bi$_2$Te$_3$}

We begin by describing the topological band structure of 
Bi$_2$Te$_3$, in order to analyze the spin projection of the Dirac electrons as well as the orbital decomposition of the states. 

For the calculation of the topological insulator Bi$_2$Te$_3$, 
we set up the atomic structure with the Atomic Simulation Environment (ASE) \cite{ASE}.
We consider 8 quintuple layers (QLs) of Bi$_2$Te$_3$, using the lattice constants \cite{Nakajima1963:JPCS}  
$a = 4.386$~\AA~and $c = 30.497$~\AA, with the atomic parameters $(u,v) = (0.4000, 0.2097)$.
In Figure \ref{Fig:Bi2Te3_geometry_bands}(a) we show the geometry of 8 QLs of Bi$_2$Se$_3$, visualized with VESTA \cite{VESTA}. The unit cell contains 40 atoms. 

The electronic structure calculations are performed by 
density functional theory (DFT)~\cite{Hohenberg1964:PRB} with {Quantum ESPRESSO}~\cite{Giannozzi2009:JPCM}.
Self-consistent calculations are performed with the $k$-point sampling of 
$30\times 30\times 1$.
The energy cutoff for the charge density is $600$~Ry, and
the kinetic energy cutoff for the wavefunctions is $70$~Ry 
We consider relativistic pseudopotentials 
with the projector augmented wave method \cite{Kresse1999:PRB} employing the 
Perdew-Burke-Ernzerhof exchange correlation functional \cite{Perdew1996:PRL}.
Dipole and van der Waals corrections \cite{Grimme2006:JCC,Barone2009:JCC,Bengtsson1999:PRB} are included to 
get correct band offsets and internal electric fields.
In order to simulate the 8 QL slab of Bi$_2$Te$_3$, we add a vacuum layer of $30$~\AA.

In Figs. \ref{Fig:Bi2Te3_geometry_bands}(c-e), we show the calculated low energy band structure, projected on the three different parts (top QL, bottom QL, bulk) of the geometry, defined in Figure \ref{Fig:Bi2Te3_geometry_bands}(a), along the $k$-path shown in Figure \ref{Fig:Bi2Te3_geometry_bands}(b). 
We find that the Dirac states, that cross the Fermi level, are localized in the top and bottom QL layer of the Bi$_2$Te$_3$. 
Due to inversion symmetry of the 8 QL structure, the Dirac states of top and bottom QL are at the same energy, but with opposite spin. 
In Figs. \ref{Fig:Bi2Te3_geometry_bands}(g-h), we show the same band structure where the color corresponds to the $s_x$, $s_y$, and $s_z$ spin expectation value, respectively. 
As the chosen $k$-path is along the $k_x$-direction, the Dirac states only have a $s_y$ spin component. 
In \ref{Fig:Bi2Te3_geometry_bands}(f) we show the spin-orbit field of the top QL Dirac bands around the $\Gamma$ point. As expected, the Dirac states show spin-momentum locking. Away from the center of the Brillouin Zone, the Dirac bands also show some trigonal warping. 
Still within the bulk gap the Dirac states acquire perpendicular
($s_z$) spin, which increases as the states get closer to the bulk bands. 

Our calculated low energy band structure agrees very well with ARPES measurements \cite{Chen2009:SC} and earlier DFT results \cite{Basak2011:PRB,Zhang2009:NP}. 
Especially the Dirac point of Bi$_2$Te$_3$ is at about $-150$~meV below the Fermi level and located within the bulk bands, see Figure \ref{Fig:Bi2Te3_geometry_bands}(c). 
In contrast, other topological insulators such as Bi$_2$Se$_3$ have the Dirac point at the Fermi level under ideal defect free conditions \cite{Zhang2009:NP}. 
However, from the experimental point of view, unintentional intrinsic  doping is present for all members of the Bi$_2$Se$_3$ topological insulator family, and the Dirac point is typically located below the Fermi level \cite{Chen2009:SC}. To compensate this effect and to bring the Dirac states to the Fermi level, the multicompositional topological insulator crystals Bi$_{2-x}$Sb$_{x}$Te$_{3-y}$Se$_y$ are 
considered \cite{Ren2011:PRB,Arakane2012:NC,Jafarpisheh2018:PRB}. 
Depending on the numbers $x$ and $y$, the defect doping can be counteracted and bulk transport can be suppressed. 
Of course, the formation of the Dirac states depends also on the number of QLs \cite{Luo2012:PRB,Liu2010:PRB,Yazyev2010:PRL,Park2010:PRL,Forster2016:PRB}, because
for thin samples the surface state wave functions still interact with each other through the bulk, such that gapless surface states are absent.

\begin{figure}[!htb]
 \includegraphics[width=.99\columnwidth]{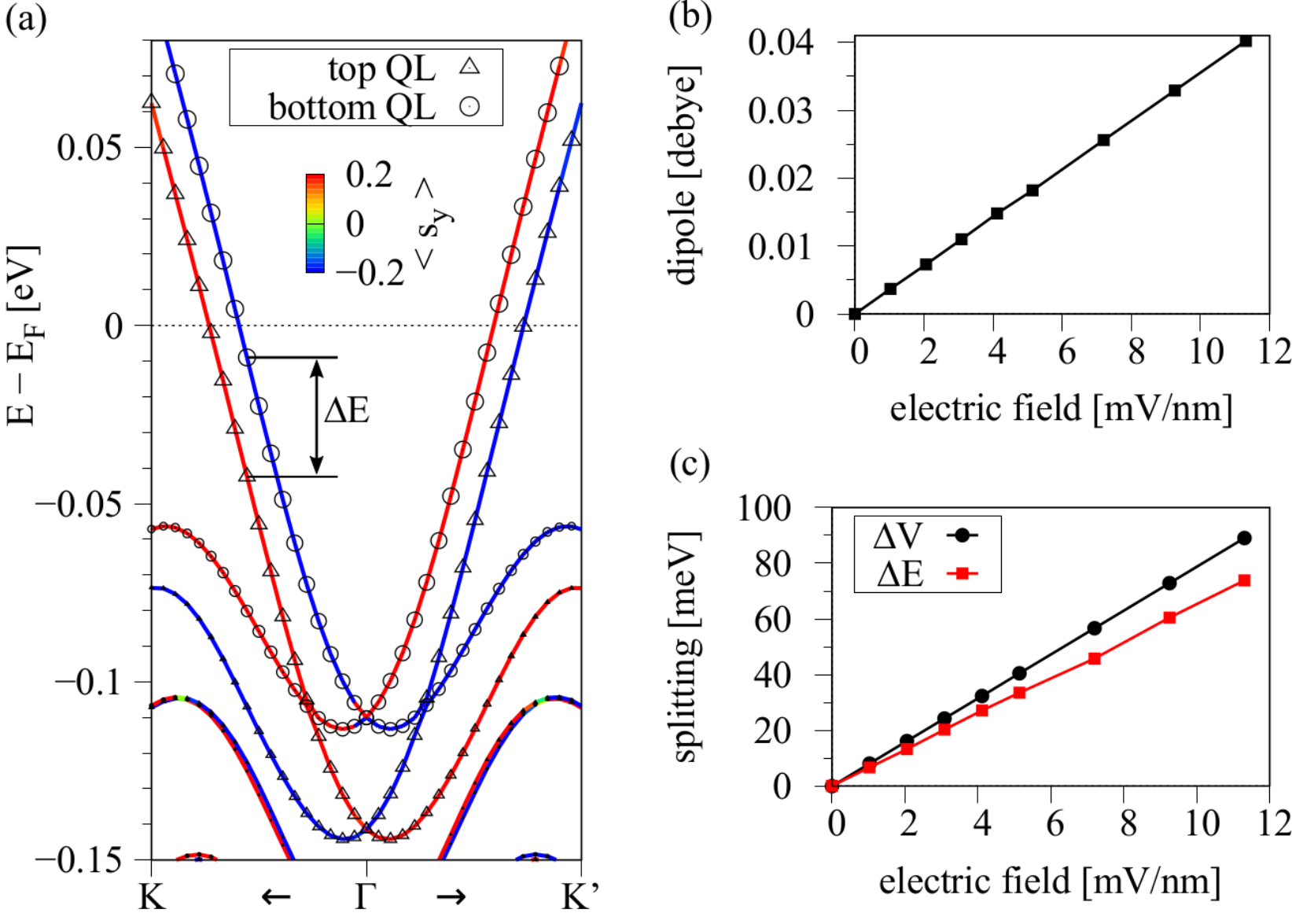}
 \caption{(a) Zoom to the calculated Dirac surface states of Bi$_2$Te$_3$ for a transverse electric field of 5~mV/nm. Color of lines corresponds to the $s_y$ spin expectation value. Open triangles (spheres) correspond to projections onto top (bottom) QL. The surface states split in energy by $\Delta E$, due to the electric field. (b) The calculated dipole and (c) potential difference $\Delta V$ and extracted energy splitting $\Delta E$ as function of the applied transverse electric field. 
 }\label{Fig:Efield_split_Bi2Te3}
\end{figure}

A possibility to break the aforementioned degeneracy of the Dirac states of top and bottom QL is by the application of a transverse electric field along the $c$-axis or through a substrate, breaking the inversion symmetry. 
Depending on the potential difference on the two sides, the Dirac states will be separated in energy \cite{Zhang2013:PRL}. 
The electric field that we apply is modeled by a sawtooth potential, and we can directly estimate
the potential energy difference $\Delta V = \mathrm{e}\times A\times d$, from the electric field amplitude $A$, the thickness of the 8 QL structure $d=78.7~\textrm{\AA}$ (the distance between the outermost Te atoms), and $\mathrm{e}$ is the charge of the electron.  
In Figure \ref{Fig:Efield_split_Bi2Te3}(a), we show a zoom to the calculated Dirac surface states of Bi$_2$Te$_3$ for a transverse electric field of 5~mV/nm.
We find that the states originating from top and bottom QL are still intact and separated 
by $\Delta E$ in energy. The dipole of the structure, see Figure \ref{Fig:Efield_split_Bi2Te3}(b), grows linearly with the applied field. 
In Figure \ref{Fig:Efield_split_Bi2Te3}(c), we compare the energy splitting $\Delta E$, extracted from the calculated band structures, with the estimated 
potential difference $\Delta V$, as function of the applied electric field. 
Both depend linearly on the applied field, as expected, but the energy splitting $\Delta E$ is smaller than the estimated potential difference $\Delta V$ for all field values. 
This can be attributed to the fact that the surface states are localized within top and bottom QL and their spatial separation is not exactly equal to the thickness $d$, as we use in the estimation for $\Delta V$. 
The splitting $\Delta E$ increases with a slope of roughly 6.5~meV per mV/nm of applied field.

 \subsection{Proximitized graphene: effective Hamiltonian with spin-orbit coupling}

In order to understand the proximity effect on the electronic band structure of graphene, we first introduce a generic phenomenological model describing Dirac states of graphene with reduced symmetry due to external effects \cite{Kochan2017:PRB,Zollner2016:PRB,Zollner2019:PRB,Zollner2019:PRB2,Gmitra2015:PRB,Gmitra2016:PRB,Sante2019:PRB,Kane2005:PRL}. The model Hamiltonian given 
in the basis $|\Psi_{\textrm{A}}, \uparrow\rangle$, 
$|\Psi_{\textrm{A}}, \downarrow\rangle$, $|\Psi_{\textrm{B}}, \uparrow\rangle$, 
and $|\Psi_{\textrm{B}}, \downarrow\rangle$ reads:
\begin{flalign}
\label{Eq:Hamiltonian}
&\mathcal{H} = \mathcal{H}_{0}+\mathcal{H}_{\Delta}+\mathcal{H}_{\textrm{I}}+\mathcal{H}_{\textrm{R}}+\mathcal{H}_{\textrm{PIA}}+E_D,\\
&\mathcal{H}_{0} = \hbar v_{\textrm{F}}(\tau k_x \sigma_x - k_y \sigma_y)\otimes s_0, \\
&\mathcal{H}_{\Delta} =\Delta \sigma_z \otimes s_0,\\
&\mathcal{H}_{\textrm{I}} = \tau (\lambda_{\textrm{I}}^\textrm{A} \sigma_{+}+\lambda_{\textrm{I}}^\textrm{B} \sigma_{-})\otimes s_z,\\
&\mathcal{H}_{\textrm{R}} = -\lambda_{\textrm{R}}(\tau \sigma_x \otimes s_y + \sigma_y \otimes s_x),\\
&\mathcal{H}_{\textrm{PIA}} = a(\lambda_{\textrm{PIA}}^\textrm{A} \sigma_{+}-\lambda_{\textrm{PIA}}^\textrm{B} 
\sigma_{-})\otimes (k_x s_y - k_y s_x). 
\end{flalign}

The first term $\mathcal{H}_{0}$ describes a gapless linear 
dispersion near Dirac points K~(K') with two-fold spin-degenerate bands. 
The parameter $v_{\textrm{F}}$ denotes the Fermi velocity, and $k_{x}$ and $k_{y}$ are the Cartesian components of 
the electron wave vector measured from $\pm$K, corresponding to the valley index $\tau = \pm 1$. 
The Pauli spin matrices are $s_i$ and $\sigma_i$ are pseudospin matrices, 
with $i = \{ 0,x,y,z \}$. We also define $\sigma_{\pm} = \frac{1}{2}(\sigma_z \pm \sigma_0)$ for shorter notation. The pristine graphene lattice constant is $a$.

When graphene is situated above a substrate, the pseudospin symmetry of graphene gets broken
and $\mathcal{H}_{\Delta}$ describes a mass term, opening a gap in the spectrum \cite{Giovannetti2007:PRB,Zhou2007:NM}. The corresponding parameter $\Delta$ is called staggered potential and models the size of the induced gap. 
Of course, the pseudospin symmetry breaking depends on the interlayer distance \cite{Zollner2019:PRB,Zollner2019:PRB2,Kou2015:C} and on the actual arrangement of graphene above the substrate's surface which can be tuned by twisting \cite{Song2018:NL,Zhang2014:PRL}.
However, the sublattice potential asymmetry is not always responsible for the gap opening. Also a Kekul\'e lattice distortion \cite{Hou2007:PRL,Giovannetti2015:PRB,Lin2017:NL,Song2018:NL,Gutierrez2016:NP}, leading to a nearest neighbor hopping asymmetry \cite{Ren2015:PRB}, and SOC, e. g., from adatoms \cite{Weeks2011:PRX,Ding2011:PRB,Jiang2012:PRL}, can open the band gap in graphene.

The Dirac bands of freestanding graphene show an intrinsic SOC of about $12~\mu$eV \cite{Gmitra2009:PRB,Sichau2019:PRL,Abdelouahed2010:PRB}. 
Due to a substrate, also the SOC in the effective graphene $p_z$ orbitals, forming the Dirac bands, can be modified. 
The term $\mathcal{H}_{\textrm{I}}$ accounts for the modification of the intrinsic SOC due to proximity effects, where $\lambda_{\textrm{I}}^\textrm{A}$ and 
$\lambda_{\textrm{I}}^\textrm{B}$ 
are the sublattice resolved intrinsic SOC parameters.

The presence of a transverse electric field (vertical to the graphene layer) 
or a substrate breaks all symmetries, that would allow to flip the orientation of the transverse $z$ axis
(inversion with respect to $z$ or mirror with respect to the $xy$-plane). 
Two additional terms arise due to this symmetry breaking, namely $\mathcal{H}_{\textrm{R}}$
 and $\mathcal{H}_{\textrm{PIA}}$.
The first term is the Rashba SOC with parameter $\lambda_{\textrm{R}}$, which describes the 
amount of space inversion asymmetry. 
The second term is the sublattice resolved pseudospin-inversion asymmetry 
(PIA) SOC Hamiltonian with parameters 
$\lambda_{\textrm{PIA}}^\textrm{A}$ and $\lambda_{\textrm{PIA}}^\textrm{B}$, 
which describe the strength of the mirror plane asymmetry. 

Finally, $E_{\textrm{D}}$ accounts for electron or hole doping of 
the Dirac bands due to external influences and we call it the Dirac point energy.

 \subsection{Graphene/topological insulator van der Waals bilayers}
\begin{figure}[!htb]
 \includegraphics[width=.99\columnwidth]{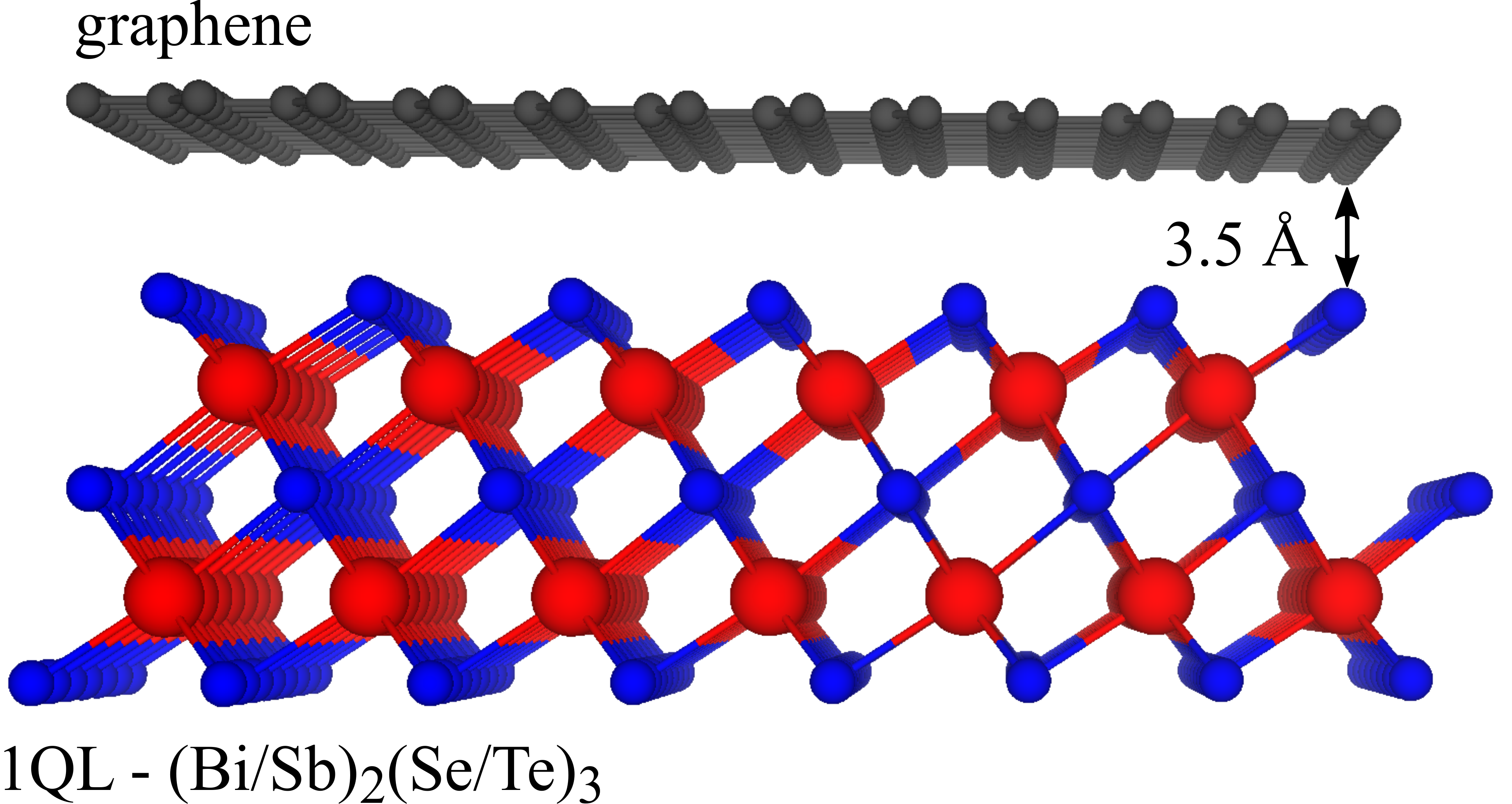}
 \caption{Geometry of graphene above one QL of (Bi/Sb)$_2$(Se/Te)$_3$.
 Different colors correspond to different atomic species, as
  in Figure \ref{Fig:Bi2Te3_geometry_bands}. 
 }\label{Fig:struct_new}
\end{figure}

We now realize, using atomistic simulations, the effective Hamiltonian introduced in the previous section by combining graphene with a single quintuplet of Bi$_2$Se$_3$, Bi$_2$Te$_2$Se, and Sb$_2$Te$_3$ topological insulators. The resulting structure is essentially a van der Waals bilayer, as we show in Figure \ref{Fig:struct_new}.

\begin{figure*}[!htb]
 \includegraphics[width=.99\textwidth]{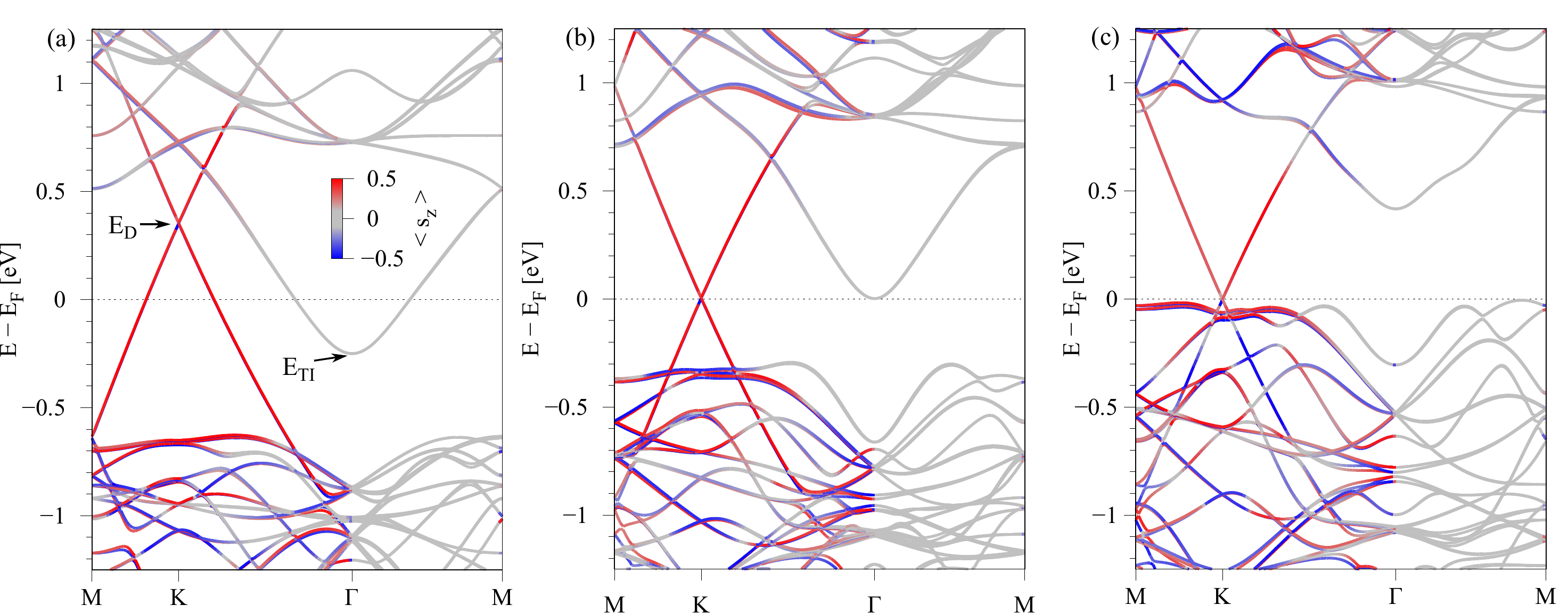}
 \caption{Calculated band structures of graphene on one QL of (a) Bi$_2$Se$_3$, (a) Bi$_2$Te$_2$Se, and (c) Sb$_2$Te$_3$. The color is the $s_z$ spin expectation value.
 In (a), we define the Dirac point energy $E_{\textrm{D}}$ and the doping energy of the topological insulator  $E_{\textrm{TI}}$.
 }\label{Fig:bandstructures_different_grp_TIs}
\end{figure*}
\begin{table*}[!htb]
\centering
\caption{\label{tab:fit_graphene_different_TIs} Fit parameters of Hamiltonian $\mathcal{H}$ 
for the graphene/topological insulator bilayers. 
The Fermi velocity $v_{\textrm{F}}$, gap parameter $\Delta$, 
 Rashba SOC parameter $\lambda_{\textrm{R}}$, 
 intrinsic SOC parameters $\lambda_{\textrm{I}}^\textrm{A}$ and $\lambda_{\textrm{I}}^\textrm{B}$, 
 and PIA SOC parameters $\lambda_{\textrm{PIA}}^\textrm{A}$ and $\lambda_{\textrm{PIA}}^\textrm{B}$.
 The Dirac point energy $E_{\textrm{D}}$, as defined in Figure \ref{Fig:bandstructures_different_grp_TIs}(a).}
\begin{tabular}{@{}l  c  c c  c  c   c  c  c@{}}
\hline
TI & $v_{\textrm{F}}/10^5 [\frac{\textrm{m}}{\textrm{s}}]$ & 
$\Delta$~[$\mu$eV]& $\lambda_{\textrm{R}}$~[meV] & $\lambda_{\textrm{I}}^\textrm{A}$~[meV] &
$\lambda_{\textrm{I}}^\textrm{B}$~[meV] & $\lambda_{\textrm{PIA}}^\textrm{A}$~[meV] & 
$\lambda_{\textrm{PIA}}^\textrm{B}$~[meV] & $E_{\textrm{D}}$ [meV] \\
\hline
Bi$_2$Se$_3$ & 8.134 & 0.6 & -0.771 & 1.142 & -1.135 & 0.465 & 0.565 & 353.2\\
Bi$_2$Te$_2$Se & 8.123 & 0.3 & -0.669 & 1.353 & -1.351 & -1.091 & -1.209 & 4.0\\
Sb$_2$Te$_3$ & 8.119 & 0.2 & -0.221 & 0.147 & -0.139 & 2.623 & 1.177 & -2.0\\
\hline
\end{tabular}
\end{table*}

For the calculation of the graphene/topological insulator bilayers
we consider a $5\times 5$ supercell of graphene
on top of a $3\times 3$ supercell of a topological insulator.
Initial atomic structures are set up with ASE \cite{ASE} and the heterostructure was visualized with VESTA \cite{VESTA}, see Figure \ref{Fig:struct_new}.
For periodic DFT calculations, we need to marginally strain the constituent layers in order to form a commensurate unit cell. Therefore, we strain the graphene lattice constant \cite{Neto2009:RMP} to $a = 2.486$~\AA~and use the lattice structure of Bi$_2$Se$_3$, according to Ref. \cite{Nakajima1963:JPCS}, extracting only 1QL of the topological insulator. 
For the other topological insulators, Bi$_2$Te$_2$Se and Sb$_2$Te$_3$, we simply replace the relevant atoms without changing the geometry. 
We consider only bilayers without relaxation, using interlayer distances of $3.5$~\AA~between the graphene layer and the QL of the topological insulator \cite{Song2018:NL,Zollner2019:PRB2}.

The first-principles calculations are performed in a similar way as for the 8QL Bi$_2$Te$_3$ structure, discussed above. 
For the bilayer structures, we use a $k$-point sampling of $9\times 9\times 1$, an energy cutoff for the charge density of $500$~Ry, and a kinetic energy cutoff for wavefunctions of $60$~Ry.
Dipole and van der Waals corrections are also included \cite{Grimme2006:JCC,Barone2009:JCC,Bengtsson1999:PRB}.
Moreover,  a vacuum layer of $24$~\AA~is added, 
to avoid interactions between periodic images in our slab geometry.

Note that for describing the electronic structure of a 3D topological insulator, the GW method is a more accurate choice, as compared to the generalized-gradient-approximation (GGA) employed in this work \cite{Aguilera2013:PRB,Forster2016:PRB,Nechaev2013:PRB,Yazyev2012:PRB}. However, as we can see in Figure \ref{Fig:Bi2Te3_geometry_bands}, the GGA also captures the main band structure features of the topological insulator and matches the ARPES measurements \cite{Chen2009:SC}.
This makes sense, since GW is usually employed to faithfully describe the orbital gap in semiconductors, while
the topological surface states are gapless (up to finite-size hybridization) and GGA is fully adequate to 
describe them and yield reliable predictions. 
Furthermore, GW calculations are computationally very demanding and inaccessible for such large heterostructure systems we consider here. 
In addition, recent GGA-based calculation results \cite{Song2018:NL} have already been successfully used for the interpretation of experimental data for graphene/topological insulator structures \cite{Jafarpisheh2018:PRB}.

In Figure \ref{Fig:bandstructures_different_grp_TIs} we show the calculated band structures of 
graphene on one QL of Bi$_2$Se$_3$, Bi$_2$Te$_2$Se, and Sb$_2$Te$_3$.
In the case of Bi$_2$Se$_3$, the Dirac point energy $E_{\textrm{D}}$ is well above the Fermi level indicating strong hole doping, similar to Refs. \cite{Song2018:NL,Chae2019:ACS}. In contrast, for the other two topological insulators, the Dirac point of graphene is located at the Fermi level. 
Since the topological insulator thickness is just 1QL, the surface states have not yet developed \cite{Zhang2010:NP,Liu2010:PRB,Yazyev2010:PRL,Park2010:PRL,Zollner2019:PRB2}. 
However, we indicate the topological insulator surface states with the energy $E_{\textrm{TI}}$ in Figure \ref{Fig:bandstructures_different_grp_TIs}(a).
By fitting the Hamiltonian $\mathcal{H}$ from the previous section to the graphene Dirac bands we can extract
several relevant orbital and spin-orbit parameters. 
In Table \ref{tab:fit_graphene_different_TIs} we show the fit parameters for the different bilayers. The accuracy of the fit is shown in the next section, where we analyze the graphene/Sb$_2$Te$_3$ case in detail. 

\begin{figure*}[!htb]
\centering
 \includegraphics[width=.99\textwidth]{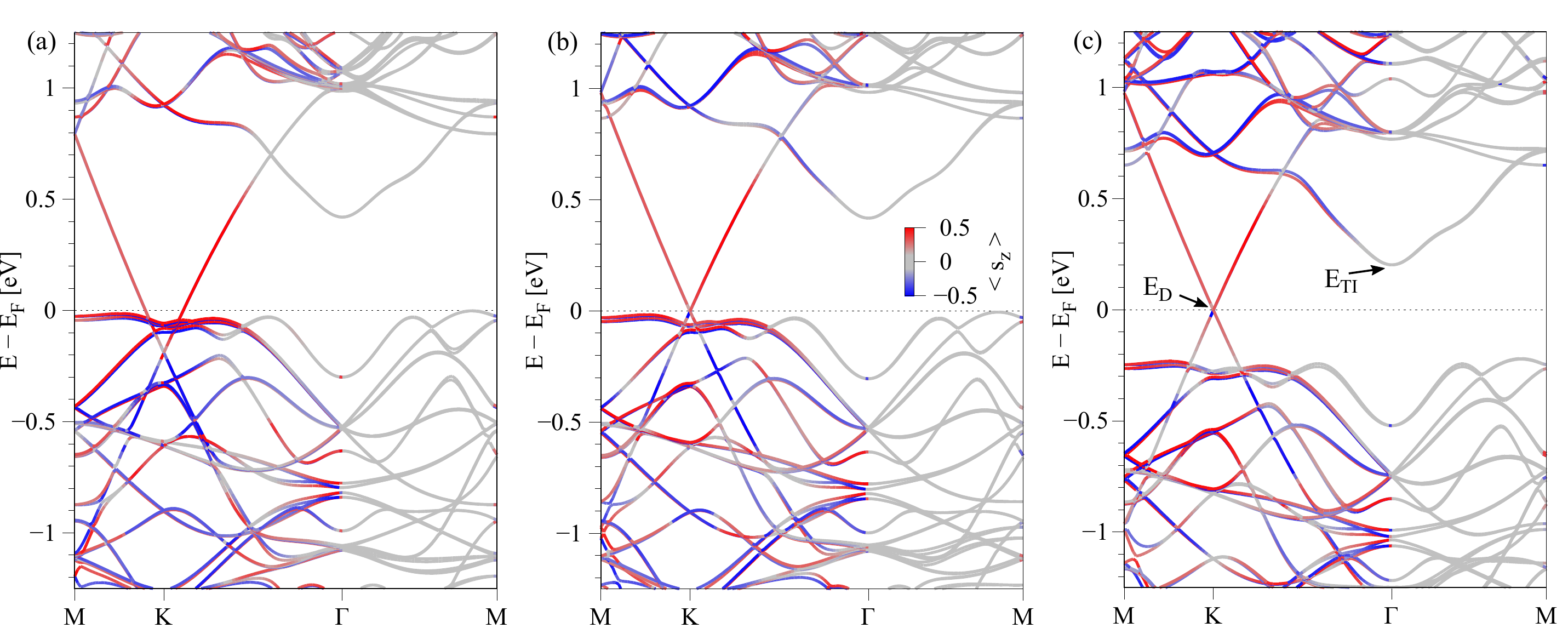}
 \caption{Calculated band structures of graphene on one QL of Sb$_2$Te$_3$ with an applied transverse electric field of (a) $-2$ V/nm, (b) $0$ V/nm, and (c) $2$ V/nm. 
 The color corresponds to the $s_z$ spin expectation value.
  In (c), we define the Dirac point energy $E_{\textrm{D}}$ and the doping energy of the topological insulator $E_{\textrm{TI}}$.
 }\label{Fig:bands_Efield_SbTe}
\end{figure*}
\begin{figure}[!htb]
 \includegraphics[width=.99\columnwidth]{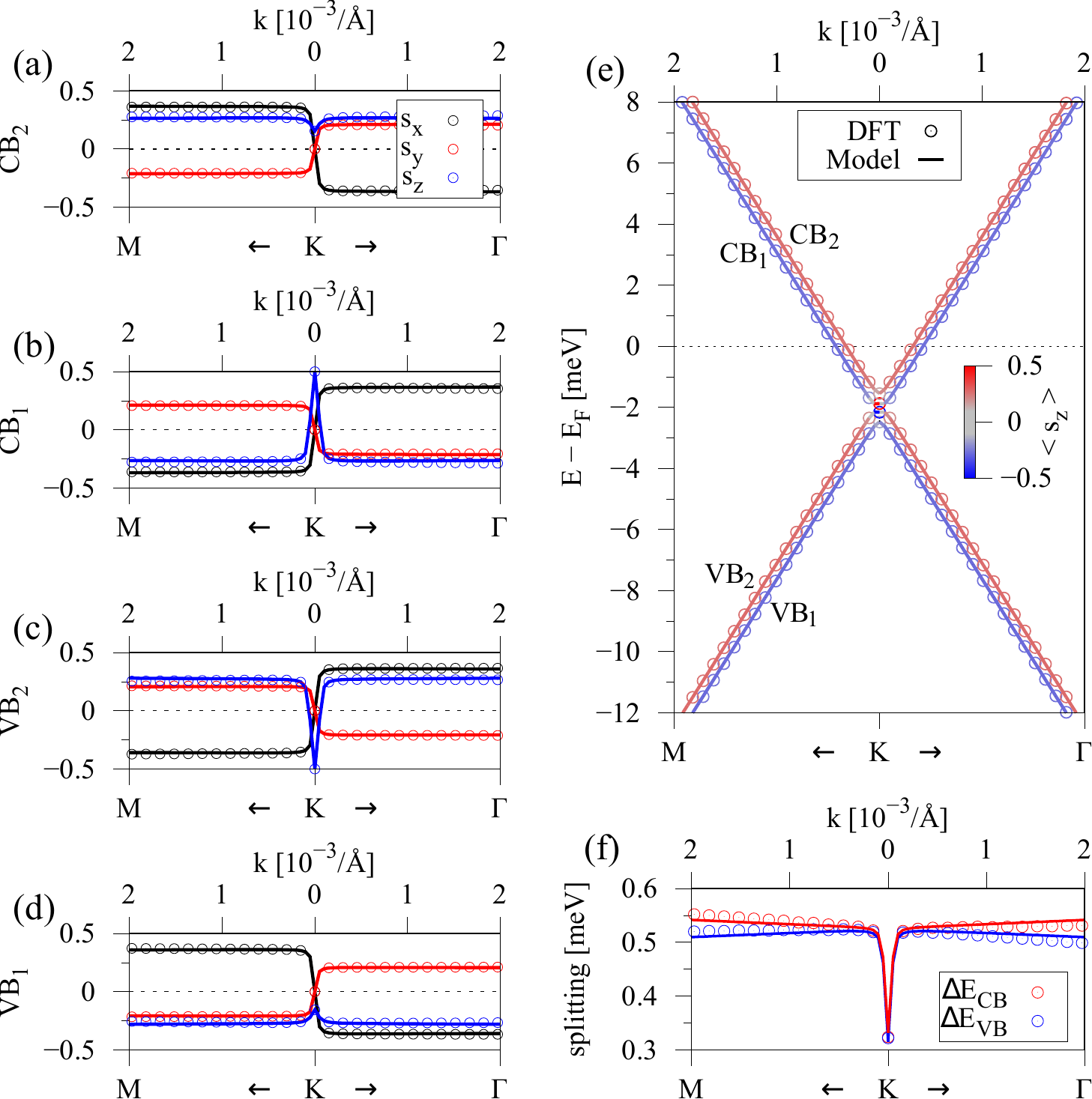}
 \caption{Calculated low energy band properties (symbols) 
 for the graphene/Sb$_2$Te$_3$ bilayer, with 
 a fit to the model Hamiltonian $\mathcal{H}$ (solid lines) for zero electric field.
 (a)-(d) The spin expectation values of the four low energy bands. 
 (e) The low energy band structure of proximitzed graphene. The color is the $s_z$ spin expectation value. 
 (f) The splitting of the valence (conduction) band in blue (red).
 }\label{Fig:spinexp_SbTe}
\end{figure}
\begin{figure}[!htb]
 \includegraphics[width=.99\columnwidth]{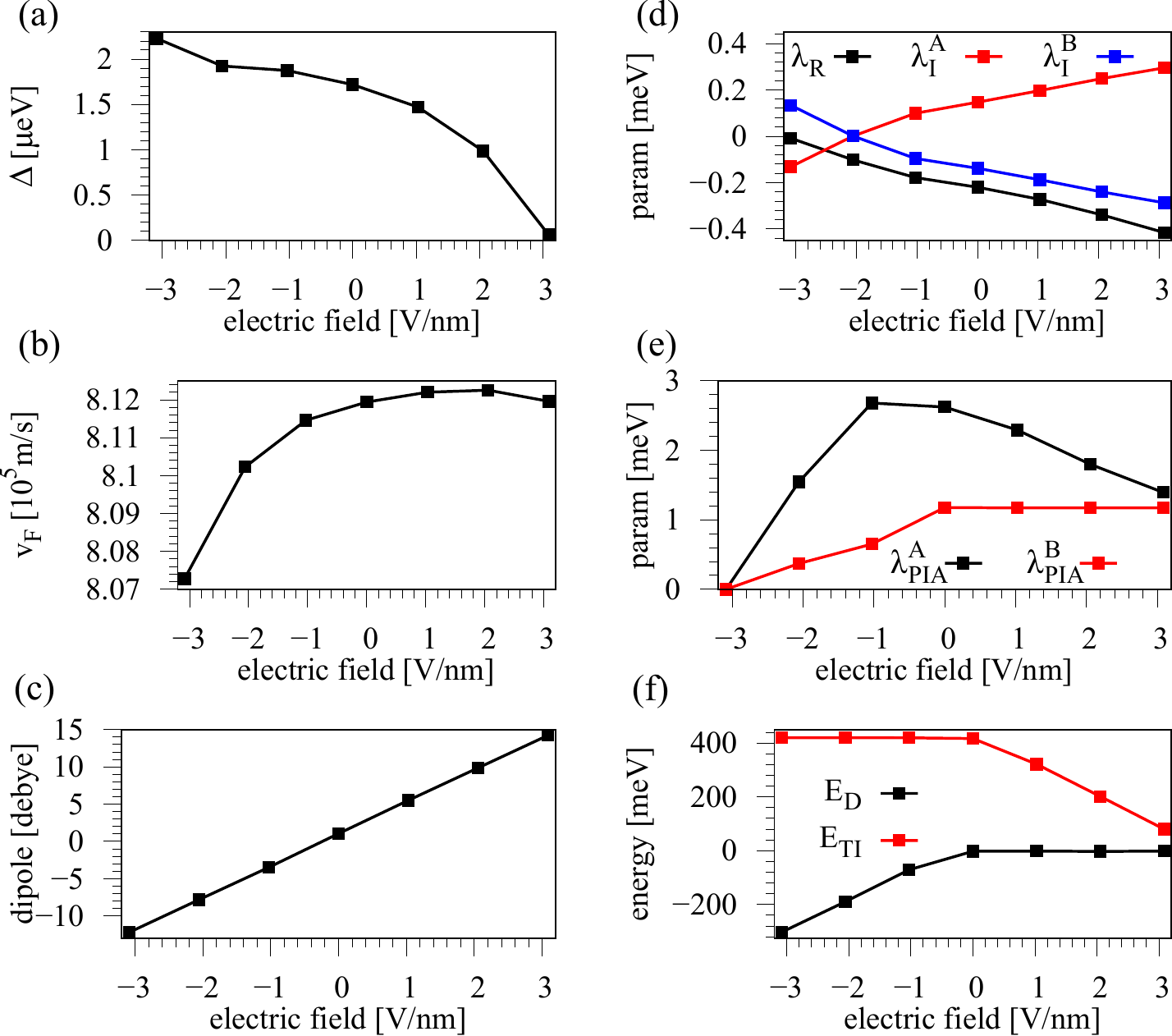}
 \caption{Fit parameters of Hamiltonian $\mathcal{H}$ 
for the graphene/Sb$_2$Te$_3$ bilayer as a function of a transverse electric field. 
(a) The gap parameter $\Delta$, (b) the Fermi velocity $v_{\textrm{F}}$, (c) the dipole of the structure, 
(d) Rashba SOC parameter $\lambda_{\textrm{R}}$, intrinsic SOC parameters $\lambda_{\textrm{I}}^\textrm{A}$ 
and $\lambda_{\textrm{I}}^\textrm{B}$, (e) PIA SOC parameters $\lambda_{\textrm{PIA}}^\textrm{A}$ and $\lambda_{\textrm{PIA}}^\textrm{B}$, and (f) the Dirac point energy $E_{\textrm{D}}$ and the doping energy of the topological insulator $E_{\textrm{TI}}$, as defined in Figure \ref{Fig:bands_Efield_SbTe}(c).
   }\label{Fig:Efield_SbTe}
\end{figure}

\begin{figure}[!htb]
\centering
 \includegraphics[width=.99\columnwidth]{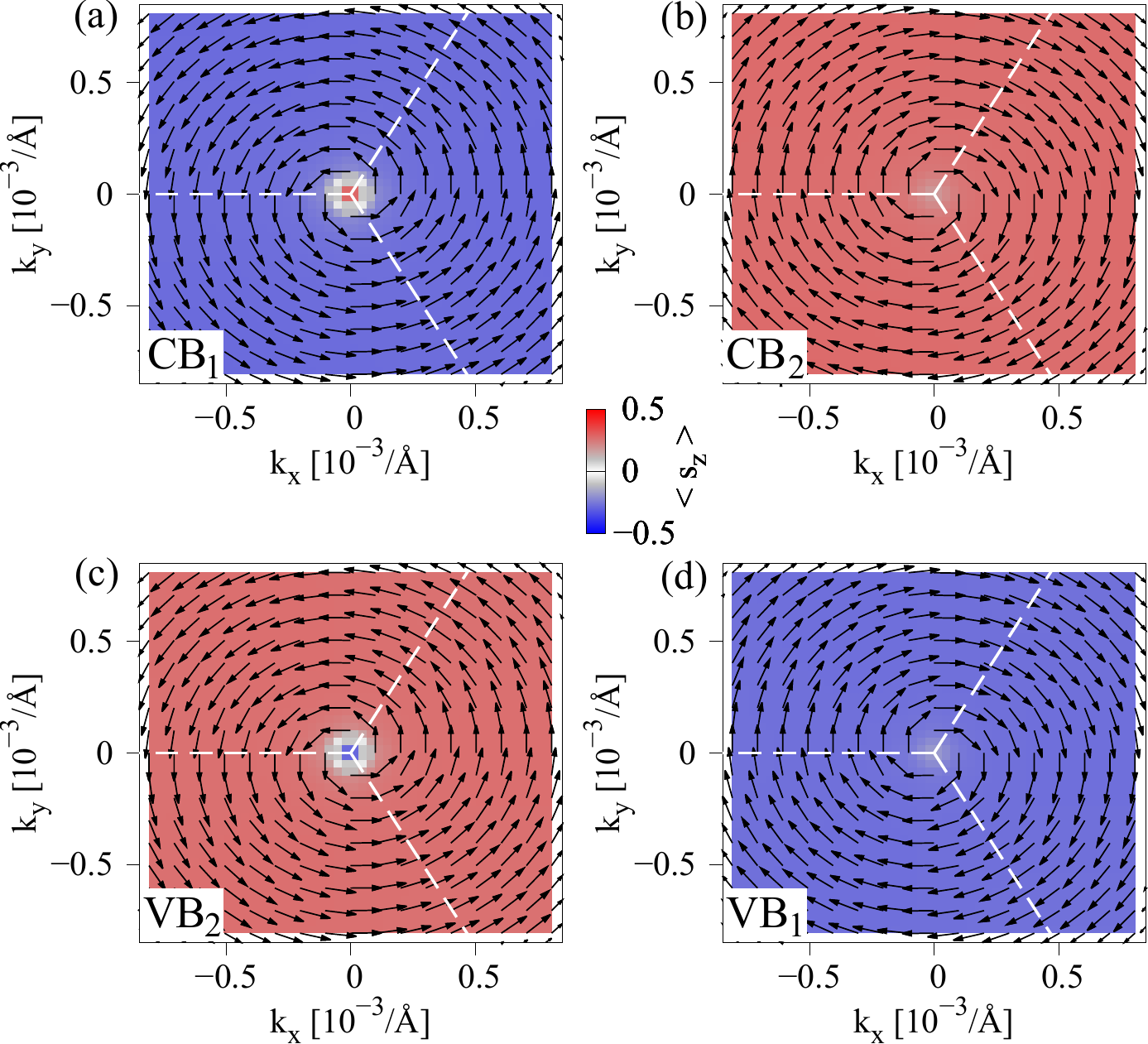}
 \caption{First-principles calculated spin-orbit fields around the K point of the bands
 (a) $\varepsilon_{1}^{\textrm{CB}}$, (b) $\varepsilon_{2}^{\textrm{CB}}$, 
 (c) $\varepsilon_{2}^{\textrm{VB}}$, and (d) $\varepsilon_{1}^{\textrm{VB}}$, 
 for the graphene/Sb$_2$Te$_3$ bilayer, 
 corresponding to the four low energy bands in Figure \ref{Fig:spinexp_SbTe}(e).
 The dashed white lines represent the edge of the Brillouin zone. }
 \label{Fig:spinfield_SbTe}
\end{figure}

The Fermi velocity is roughly independent of the topological insulator substrate. The sublattice symmetry breaking of graphene due to the topological insulator, described by the 
staggered mass parameter $\Delta$, is tiny and almost negligible compared to the other parameters.
Consequently, the potential asymmetry of the graphene sublattices in our investigated structure is small. However, by twisting the layers \cite{Song2018:NL,Zhang2014:PRL} or by decreasing the interlayer distance between graphene and the topological insulator surface \cite{Zollner2019:PRB,Zollner2019:PRB2,Kou2015:C}, the gap in graphene's spectrum can be enhanced.

Interestingly, the intrinsic SOC parameters $\lambda_{\textrm{I}}^\textrm{A}$ and $\lambda_{\textrm{I}}^\textrm{B}$ are almost equal in magnitude but opposite in sign for all the studied bilayers. 
Such a valley-Zeeman type SOC, i. e. $\lambda_{\textrm{I}}^\textrm{A} \approx -\lambda_{\textrm{I}}^\textrm{B}$, can lead to giant spin-relaxation anisotropies in graphene \cite{Cummings2017:PRL,Gmitra2016:PRB,Song2018:NL}.
A more detailed analysis of the graphene/Bi$_2$Se$_3$ and graphene/Bi$_2$Te$_2$Se cases is given in Refs.  \cite{Zollner2019:PRB2,Song2018:NL}.
More precisely, depending on the twist angle and the exact interface of graphene and the topological insulator, a giant spin relaxation anisotropy can be present \cite{Song2018:NL}. Additional QLs of the topological insulator are necessary for the surface states to form, but will have a minor extra impact on graphene's band structure, since proximity effects are short-ranged. In contrast, two very efficient tunability knobs for band offsets and proximity SOC are the interlayer distance and a transverse electric field \cite{Zollner2019:PRB2}.

Different to monolayer graphene, bilayer graphene shows a giant band gap, due to the intrinsic dipole present in heterostructures with Bi$_2$Se$_3$.
Moreover, the resulting proximity band structure of bilayer graphene can be tuned by gating and a spin-orbit valve can be realized \cite{Zollner2019:PRB2,Gmitra2017:PRL}.

\subsection{\texorpdfstring{Sb$_2$Te$_3$ substrate}{}}

The proximity effect in graphene due to Sb$_2$Te$_3$
has not yet been systematically studied. Below we provide 
both DFT results and phenomenological descriptions for these
bilayers. 

In Figure \ref{Fig:bands_Efield_SbTe}(b) we show the calculated band structure for the graphene/Sb$_2$Te$_3$ heterostructure, 
in the absence of a transverse electric field applied across the bilayer structure. 
We find that the Dirac point of graphene, as well as the band edge originating from the topological insulator is located at the Fermi level. 
The overall band structure is comparable to ARPES measurements of graphene on a thick Sb$_2$Te$_3$ substrate, showing the coexistence of both Dirac cones near the bulk Sb$_2$Te$_3$ valence band edge \cite{Bian2016:2DM}.

When a negative transverse electric field of $-2$~V/nm is applied across the bilayer, see Figure \ref{Fig:bands_Efield_SbTe}(a), graphene gets electron doped and the Dirac point shifts to about $-200$~meV below the Fermi level. 
The bands of the topological insulator do not shift in energy, compared to the zero field case. 
In contrast, when a positive electric field of $2$~V/nm is applied, see Figure \ref{Fig:bands_Efield_SbTe}(c), the graphene bands do not shift in energy, while the bands of the topological insulator do.
In Figure \ref{Fig:bands_Efield_SbTe}(c), we also label the \textit{doping energy} of the topological insulator with $E_{\textrm{TI}}$, as these bands would correspond to the topological surface states in few QL structures. 
In Figure \ref{Fig:spinexp_SbTe}, we show the low energy band properties of the graphene Dirac states, fitted to the model Hamiltonian, for zero electric field. The model agrees perfectly
with the DFT calculated band structure, capturing also the spin expectation values and band splittings, using the parameters summarized in Table \ref{tab:fit_graphene_different_TIs} for the Sb$_2$Te$_3$ substrate. 

In Figure \ref{Fig:Efield_SbTe} we summarize the evolution of the fit parameters as function of a transverse electric field, applied across the bilayer. 
Most interesting are the intrinsic SOC parameters, which can be tuned from positive to negative values, but always of valley-Zeeman type. The Rashba and PIA SOC parameters are also strongly changing with the applied field and can be even tuned to zero. 
The resulting spin-orbit fields of the Dirac bands are due to a competition of Rashba and PIA SOC favoring an in-plane spin texture, and the intrinsic SOCs favoring an out-of-plane texture. 
Due to tunability of these parameters with the electric field, we have a potential knob to tune the spin-orbit fields, as well as the magnitude of the proximity-induced SOC. 
The spin-orbit fields of the four Dirac bands, as labeled in Figure \ref{Fig:spinexp_SbTe}(e), are shown in Figure \ref{Fig:spinfield_SbTe} for the zero field case. 
We can see that bands show very pronounced Rashba spin-orbit fields. For example, the first conduction band (CB$_1$) shows counter-clockwise, while the second conduction band (CB$_2$) shows a clockwise rotating spin-orbit field, both also with a significant and opposite out-of-plane spin component. 

Recently, a gate-tunable spin-galvanic effect has been shown experimentally in graphene/topological insulator bilayers \cite{Khokhriakov2019:arxiv}. 
More precisely, they demonstrate an efficient spin-charge conversion at room temperature in graphene/(Bi$_{0.15}$Sb$_{0.85}$)$_2$Te$_3$ heterostructures, which should be well comparable to our graphene/Sb$_2$Te$_3$ bilayers. 
Especially the electric field results in Figure \ref{Fig:Efield_SbTe} can be used to explain their gate-tunability of the conversion efficiency, due to tunable proximity SOC. 

Based on the above results, we can conclude that for device applications, only a thin (1--2 QLs) topological insulator is sufficient to fully exploit it's proximity effect on graphene. A thicker topological insulator is necessary for the Dirac surface states to form, allowing to simultaneously study two types of Dirac electrons, with very different spin-orbit fields. An electric field can be used to tune both, the surface states of the topological insulator and the proximity SOC in graphene. 
The magnitude of proximity SOC and band offsets depend on the topological insulator crystal. Consequently, a multicompositional material Bi$_{2-x}$Sb$_{x}$Te$_{3-y}$Se$_y$ might be the best choice for applications, since proximity effects can be maximized with energetically aligned Dirac states. Especially the mentioned gate tunable spin-charge conversion is important for novel spin-orbit technology, without the need of ferromagnets.

\section{Summary}

We have reviewed the basic properties of the topological insulator Bi$_2$Te$_3$
and find gate tunable energy splitting of Dirac states, which results from the potential difference in the surface states. The energy splitting increases linearly with a slope of about 6.5~meV per mV/nm of applied field, which can be experimentally verified.
Additionally, we have reported original results for graphene/Sb$_2$Te$_3$ bilayers in the context of related
graphene heterostructures with Bi$_2$Se$_3$ and Bi$_2$Te$_2$Se. 
We find that the position of the graphene Dirac point strongly depends on the substrate, when considering a single quintuplet of Bi$_2$Se$_3$, Bi$_2$Te$_2$Se, or Sb$_2$Te$_3$.
We quantify the proximity SOC by fitting a symmetry-derived low energy graphene Hamiltonian to the DFT simulated band structure. 
The overall results are similar for all different topological insulators; we find a strongly enhanced SOC in graphene, which is of the valley-Zeeman type. 
The effective model and fitted parameters provide
realistic foundations for phenomenological modeling of especially spin transport, and for interpreting future experiments on such structures.

From the detailed analysis of the graphene/Sb$_2$Te$_3$ case, we find also a strongly gate tunable proximity SOC and doping level.
We show that by tuning the gate field the graphene Dirac point can be well isolated from the valence band of the topological insulator, and the spin-orbit parameters can change sign as a function of the electric field. Remarkably, for all the investigated electric fields the intrinsic SOC induced in graphene remains of the valley Zeeman type, although the corresponding parameters change sign (simultaneously) at around the fields of about $-2$ V/nm. For this particular field value the Rashba coupling is predicted to dominate the spin properties. Our results regarding the electric field tunability of the proximity SOC strength is important to interpret recent gate-tunable spin-charge conversion experiments.

As outlook, it will be important to make a systematic investigation of twisted bilayers of graphene and topological insulator quintuplets, to demonstrate further tunability of the proximity induced phenomena in the two important materials.

\begin{acknowledgement}
This work was supported by the Deutsche Forschungsgemeinschaft 
(DFG, German Research Foundation) SPP 1666.
\end{acknowledgement}

\section*{Conflict of Interest}
The authors declare no financial or commercial conflicts of interests.

\section*{Table of Contents}
Proximity effects are a vital route to modify the electronic states of neighboring materials. Three-dimensional topological insulators, such as Bi$_2$Te$_2$Se, host topological surface states, due to their strong spin-orbit coupling. When graphene is placed on the surface of a topological insulator, giant spin-orbit coupling is induced by the proximity effect, enabling interesting novel electronic properties of its Dirac electrons.

\begin{figure}[!htb]
\centering
 \includegraphics[width=.99\columnwidth]{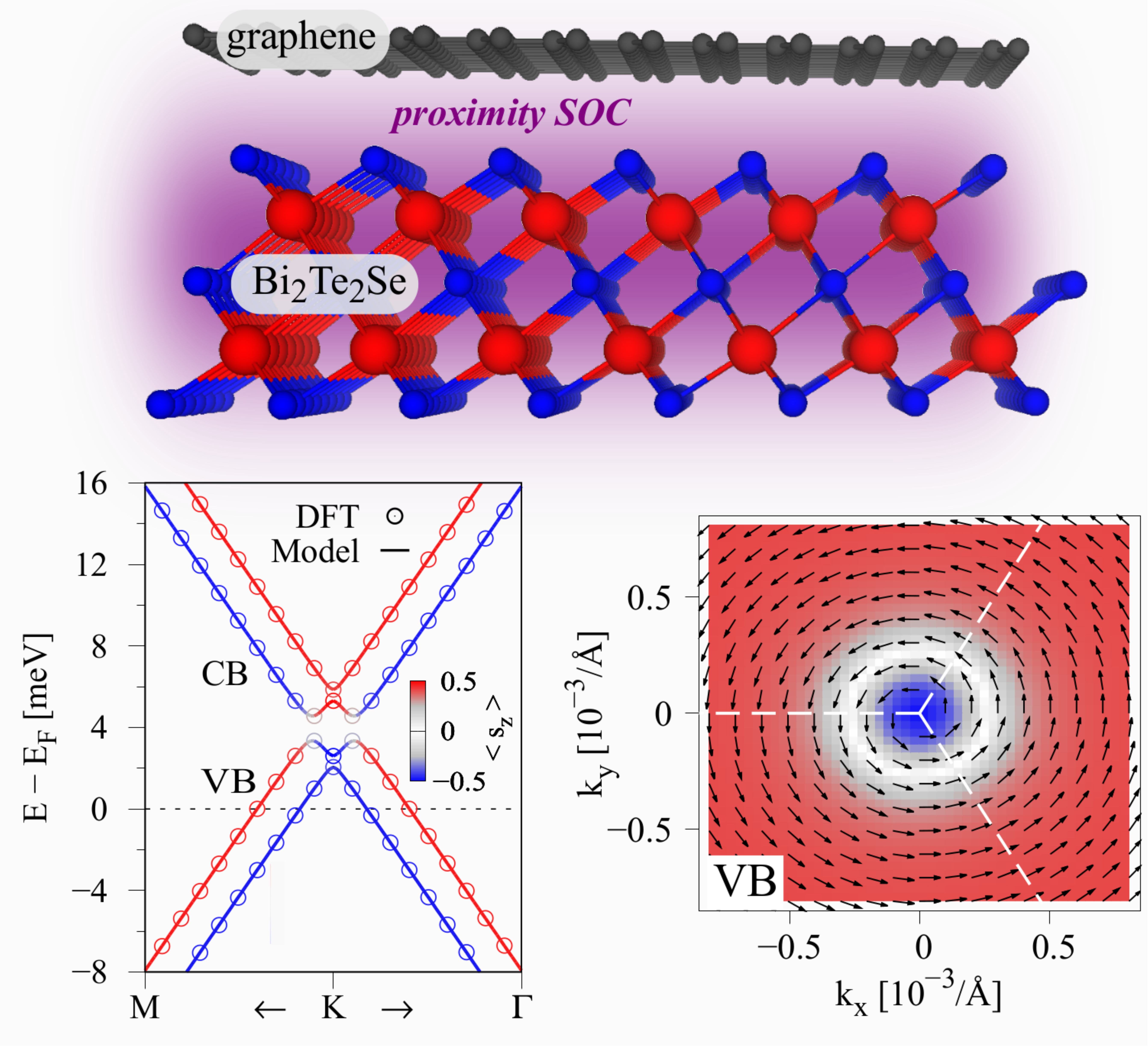}
 \caption{Table of Contents Graphic}
 \label{Fig:TOC}
\end{figure}

\bibliographystyle{pss}
\bibliography{paper}

\end{document}